\newcommand{\Teff}{\ensuremath{T_{\mathrm{eff}}}}
\newcommand{\pun}[1]{\mbox{\rm\,#1}} 
\newcommand{\moh}{\ensuremath{[\mathrm{M/H}]}}
\documentclass[structabstract]{aa}  
\usepackage{graphicx}
\usepackage{txfonts}
\usepackage{supertabular}
\usepackage{longtable}
\begin{document}
   \title{Convective line shifts for the Gaia RVS from the CIFIST 3D model atmosphere grid 
   \thanks{The authors acknowledge the Texas Advanced Computing Center (TACC) at
The University of Texas at Austin for providing HPC resources that have
contributed to the research results reported within this paper (www.tacc.utexas.edu). 
The model spectra discussed in this paper are available in electronic form
at the CDS via anonymous ftp to cdsarc.u-strasbg.fr (130.79.128.5)
or via http://cdsweb.u-strasbg.fr/cgi-bin/qcat?J/A+A/}}

   \author{C. Allende Prieto
          \inst{1}
          \fnmsep\inst{2}
          \and
          L. Koesterke
          \inst{3}
          \and
          H.-G. Ludwig
          \inst{4}
          \and
          B. Freytag
          \inst{5}
          \and
          E. Caffau
          \inst{4}
          }

   \institute{Instituto de Astrof\'{\i}sica de Canarias, 
   				E-38205  La Laguna, Tenerife, Spain 
         \and
             Departamento de Astrof\'{\i}sica, Universidad de La Laguna, 
                E-38206 La Laguna, Tenerife, Spain\\
             \email{callende@iac.es}
         \and
             Texas Advanced Computing Center, 
             University of Texas, Austin, TX 78759, USA\\
             \email{lars@tacc.utexas.edu}             
         \and
         	 Zentrum f\"ur Astronomie der Universit\"at Heidelberg,
			Landessternwarte, K\"onigstuhl 12, D-69117 Heidelberg, Germany\\
             \email{H.Ludwig@lsw.uni-heidelberg.de, E.Caffau@lsw.uni-heidelberg.de}
		 \and
Centre de Recherche Astrophysique de Lyon,
UMR 5574: CNRS, Universit\'e de Lyon,
\'Ecole Normale Sup\'erieure de Lyon,
46 all\'ee d'Italie, F-69364 Lyon Cedex 07, France \\
             \email{Bernd.Freytag@ens-lyon.fr}
             }

   \date{Submitted July 20, 2012}

 
  \abstract
   {To derive space velocities of stars along the line of sight from 
   wavelength shifts in stellar spectra requires accounting for a number
   of second-order effects. For most stars, gravitational redshifts, 
   convective blueshifts, and transverse stellar motion are the 
   dominant contributors.}
   {We provide theoretical corrections for the net   
   velocity shifts due to convection expected for the measurements from 
   the Gaia Radial Velocity Spectrometer (RVS).}
   {We used a set of three-dimensional time-dependent simulations of
   stellar surface convection computed with CO5BOLD to calculate 
   spectra of late-type stars in the Gaia RVS range
   and to infer the net velocity offset that convective motions
   will induce in radial velocities derived by cross-correlation.}
   {The net velocity shifts derived by cross-correlation depend both on 
    the wavelength range and  spectral resolution of the observations. 
   Convective shifts for Gaia RVS observations are less than 0.1
   km s$^{-1}$ for late-K-type stars, and they increase with stellar mass, 
   reaching about 0.3 km s$^{-1}$ or more for early F-type dwarfs. 
   This tendency is the result of an increase with effective temperature
in both temperature and velocity fluctuations in the line-forming region. 
Our simulations also indicate that the 
net RVS convective shifts can be positive 
   (i.e. redshifts) in some cases. Overall, the blueshifts weaken slightly 
   with increasing surface gravity,  and are enhanced at low metallicity. Gravitational 
  redshifts amount up to 0.7 km s$^{-1}$ and dominate  
  convective blueshifts for dwarfs, but become much weaker for giants. 
   }
   {}


   \maketitle
%

\section{Introduction}

Stellar atmospheres are not in hydrostatic equilibrium. 
Convection, magnetic fields, rotation, oscillations, and other phenomena
induce time and spatially changing structure on the surfaces
of stars. In the case of late-type stars, surface convection 
is responsible for the granulation pattern  directly 
 observed in high-resolution pictures of the Sun.  

Granulation
broadens absorption line profiles in observed solar 
spectra, making them asymmetric and introducing 
a net blueshift in their central wavelengths.
Even though the typical granular velocities reach several 
kilometers per second, the net blueshift observed in 
spatially-averaged spectra of solar photospheric 
lines rarely exceeds 1 km s$^{-1}$.  
In solar-type stars, the photosphere is usually located above 
the convectively unstable zone and the velocity field weakens 
with height. This causes the net convective shift 
to be strongest for  lines formed in deep layers,
progressively weakening and then disappearing for  lines
formed higher up in the photosphere (Allende Prieto 
\& Garc\'{\i}a L\'opez 1998; Pierce \& Lopresto 2000; 
Ram\'irez, Allende Prieto \& Lambert 2008; Gray 2009).  
     
Stellar spectra are often used to measure the velocities of stars projected
along the line of sight.  The interpretation of the measured line wavelength
shifts in terms of Doppler shifts purely associated with bulk stellar motion
results in biased radial velocities owing to the presence of photospheric
velocity fields. Furthermore, motions in stellar atmospheres are not the only
mechanism that can bias radial velocity measurements, and a number of other
effects, most notably gravitational redshifts (e.g., Pasquini et al. 2011),
frustrate the direct derivation of the line-of-sight component of the velocity
of a star's center of mass from the spectrum. Correcting for all these effects
involves knowledge that is usually uncertain and, in most cases, model
dependent, and therefore the IAU recommends using the term
'radial-velocity measure' to refer to the velocity naively inferred from the
spectral shifts under classical Newtonian mechanics, ignoring the corrections
mentioned earlier (Lindegren \& Dravins 2003).

The ESA mission Gaia is scheduled for launch in late 2013 with the
purpose of measuring proper motions, parallaxes, and spectrophotometry
for some 10$^9$ stars down to $V\sim 20$ mag, and spectra for
about 10$^8$ stars down to $V\sim 17$ mag (de Bruijne et al. 2009; 
Lindegren 2010). The typical uncertainty
of the radial-velocity measures provided by Radial Velocity
Spectrometer (RVS) onboard Gaia (Katz et al. 2004; Wilkinson et al. 2005) 
will  typically be in the range 1--20 km s$^{-1}$, but taking the average 
for large  groups of stars will   largely reduce the uncertainties, 
and the systematic errors discussed above can become dominant if neglected. 
More important, the RVS does not have any onboard calibration source 
to rely on for wavelength calibration, and systematic effects need 
to be considered, since wavelength calibration is based on 
a subset of the very same stellar spectra obtained by RVS (Katz et al. 2011).

As nature  favors the formation of stars with low masses, 
we focus our attention on late-type stars, for which systematic 
errors affecting the translation 
from spectral shifts to bulk
velocities  are mainly gravitational redshifts,
convective shifts, and the effect of the transversal motion on the
relativistic Doppler effect. With the parallaxes measured by Gaia,
through the comparison of colors and absolute brightness with
stellar evolution theory, it will be possible to estimate masses and
radii to calculate  gravitational redshifts to better than 10 \%
(Allende Prieto \& Lambert 1999), or some 50 m s$^{-1}$ for a solar-like
star. Accurate proper motions and parallaxes will also provide accurate
transversal motions. With the exception of the Sun, it is very 
difficult to disentangle 
convective shifts from the bulk stellar motion, 
and therefore it becomes necessary to rely on models.

We have shown
that modern three-dimensional hydrodynamical models of the solar surface
can accurately predict the convective shifts of individual weak and
moderate-strength photospheric iron lines to within 
70 m s$^{-1}$ (Allende Prieto et al. 2009).  The
same models, however, predict net redshifts, which are not observed, for
very strong iron lines (Allende Prieto et al. 2009; see also Asplund et
al. 2000).  Convective shifts need to be evaluated for specific spectral
windows used in the derivation of radial velocities, ensuring that
models are appropriate for those windows.  

In this paper, we focus our
attention on the observations to be obtained by the Gaia RVS and
predict convective wavelength shifts for stars of spectral types M
through A ($3500<T_{\rm eff} <7000$ K) from a collection of
hydrodynamical simulations presented by Ludwig et al. (2009). 
As such corrections are specifically derived for the Gaia RVS, and in 
particular for radial velocity measures derived by cross-correlation of
RVS spectra with template spectra calculated from classical (hydrostatic) 
model atmospheres, we will refer to them, indistinctively, as net convective 
shifts or effective convective shifts for RVS spectra.

Section 2 describes the hydrodynamical simulations.
Section 3 is devoted to the spectral synthesis calculations, and \S
4 describes the results. In \S 5 we comment briefly on 
the size of the gravitational
shifts expected for the same stars previously modeled, and Section 6 
presents a summary of our findings.

\section{Model atmospheres}
\label{models}

Simulations of stellar surface convection for 83 combinations of surface
gravity, entropy flux at the bottom of the atmosphere, 
and  chemical composition were computed using the
code CO$^5$BOLD (Freytag, Steffen \& Dorch 2002; Wedemeyer et al. 2004;
Freytag et al. 2012).
The code solves the equations of hydrodynamics in a Cartesian grid with
periodic horizontal boundary conditions and accounting for the effect
of the  radiation field on the energy balance. 

The models typically have a resolution of $140\times140\times150$ (X-Y-Z) grid
points, corresponding to physical sizes significantly larger than 
the typical size of granules, and range from a few Mm for dwarfs to
thousands of Mm for giant stars.
The radiation field applying a multi-group technique
(Nordlund 1982) using 5 groups for models of solar and 6 groups for models 
of subsolar metallicity. 
The number of groups was chosen by trial-and-error to ensure an acceptable
representation of the radiative cooling or heating effects.
This was done primarily by looking at solar-like stars. It turned out,
however, that also giants, as well as hotter and cooler dwarfs could be
treated in the same way obtaining similar accuracies.  
It was found that at subsolar metallicities the treatment of the radiative 
transfer in the continuum is more critical, so that an extra group was introduced.

We believe that the typically employed five or six groups 
are sufficient for
obtaining accurate line shifts, but this is difficult to 
demonstrate in general since only for few cases are models with 
more groups available. However,
the  line shifts hinge very much on the velocity field which is driven in
deeper layers, where the wavelength dependence of the radiative transfer
becomes of secondary importance. In the few models employing 12 groups
that are available, changes in the level of the temperature fluctuations
are not large. We think that rather spatial resolution and sampling 
of the time series dominate the uncertainties in the line shifts.

The equation of state takes into account H and He
ionization and H$_2$ formation, while the opacities are calculated with the
Uppsala package (Gustafsson et al. 2008). The physical size of the
computational box is the same for models with the same surface temperature and
gravity. The upper boundary of the models is open, and extended beyond a
Rosseland optical depth of 10$^{-6}$, while 
the lower boundary  is well into optically
thick, convectively unstable layers. Each
simulation was run until fully relaxed.  Chemical abundances are from Grevesse
\& Sauval (1998), with the exception of CNO, which are updated following
Asplund (2005).  More details are provided by Ludwig et al. (2009).

For each simulation, a set of uncorrelated
snapshots was selected for spectral synthesis.
The number of snapshots and the total time span of the simulations varied from
star to star, according to the granular turn-over time scale for each
model. The snapshots were hand chosen in order to be representative of the
whole simulated time series, avoiding statistical anomalies.
We found from tests that about 20 uncorrelated snapshots 
are enough to represent a model, but in some cases a smaller
number was adequate.

In order to determine the net convective shift predicted for each 3D model, we
calculated a 1D model atmosphere with the same effective temperature, surface
gravity and metallicity. The effective temperature for each simulation was
determined from the temporally and spatially averaged emergent bolometric
flux.  The 1D models were calculated with the LHD package (Caffau \& Ludwig
2007), and share the basic ingredients (most importantly the opacities and
equation of state) with the 3D simulations.  A second comparison set of 1D
models was derived by linear interpolation of the ODFNEW model grid by
Castelli \& Kurucz\footnote{kurucz.harvard.edu} (2004), based on the solar
abundances of Grevesse \& Sauval (1998).

\begin{figure*}[t!]
\footnotesize
\centering
\includegraphics[width=13cm,angle=90]{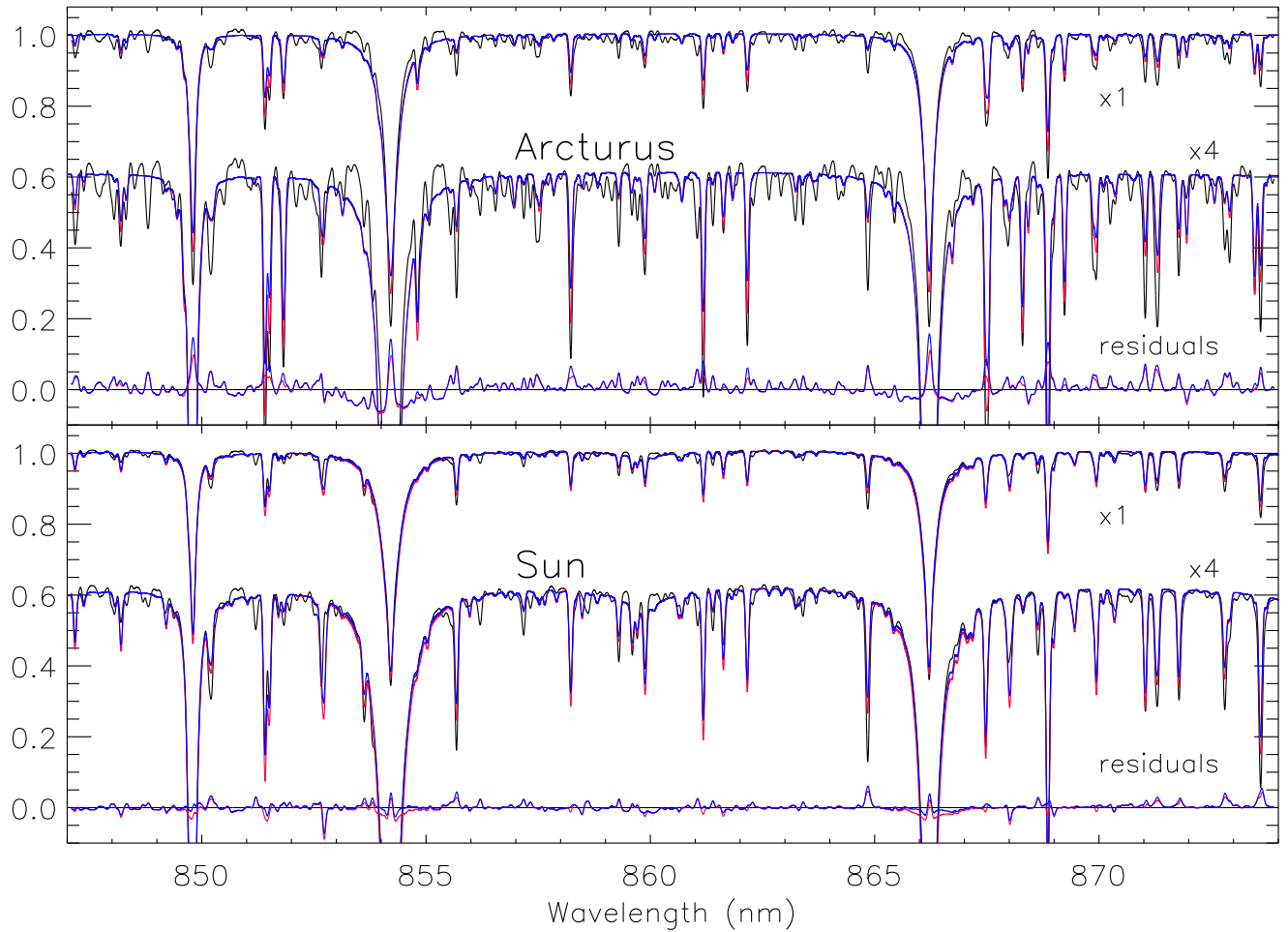}
\caption{Observed (black line) and synthetic (1D in red, 3D in blue) 
spectra for Arcturus and the Sun -- all data smoothed with a Gaussian kernel
to have a resolving power of about 11,500 as appropriate for the Gaia RVS. 
Both the normalized spectra  
 and the residuals are shown for each star.
A microturbulence of 1.5 km s$^{-1}$ was adopted
for the 1D models, which were calculated with the LHD package.}
\label{f1}%
\end{figure*}

\section{Spectral synthesis}
\label{synthesis}

\subsection{3D spectral synthesis}
\label{3ds}

We  computed synthetic spectra covering the Gaia RVS 
wavelength range (847--874 nm) with wavelength steps 
equivalent to $\leq 0.1$ km s$^{-1}$ (or a resolving power 
$R \equiv \lambda/\delta\lambda \sim 10^6$). 
The calculations were 
carried out with the 3D synthesis code ASS$\epsilon$T 
(Koesterke 2009; Koesterke, Allende Prieto \& Lambert 2008). 
The reference solar abundances are from 
Asplund, Grevesse \& Sauval (2005), 
and while temperature, density and velocity fields 
were taken from the model atmospheres, the electron density was recalculated 
with an equation of state including the first 99 elements
in the periodic table and 338 molecules (Tsuji 1973, with partition
functions from Irwin 1981). 

The number of snapshots used varied between 8 and 21, 
but in the vast majority of cases was about 20, and the total time
covered ranged between 0.6 and 19500 hours, spanning at least 10 times the typical lifetimes of convective cells. 
The  number of pixels considered for
each snapshot was reduced by taking one out of three points in the X- and
Y-directions, and at least the uppermost 90 vertical points. 
These choices are based on previous experience and limited testing
with the set of models used in this paper. These tests indicate that 
dropping spatial points would lead only to differences at the level of 
about 0.2\% for a typical strong Fe I line. For 
the continuum and weak lines the differences tend to be much smaller
than that.

The adopted solar abundances are very similar to those
used in the 3D simulations. The small existing inconsistencies are 
expected to have a negligible impact in the predicted convective shifts.
Continuous  absorption from H and H$^-$  is included.
Line absorption is included in detail from the atomic and molecular
files compiled by Kurucz. Opacities are precomputed in a 
temperature-density grid bracketing all the snapshots 
with steps of 250 K and 0.25 dex, 
and subsequently derived for the grid points using cubic B\'ezier 
interpolation. 

For each individual snapshot a spectrum is derived in 2 steps. First a
background radiation field is calculated. This is used for the
scattering term in the subsequent spectrum synthesis. Note that actually
2 spectra are calculated for each snapshot. The second one, which is
derived from opacity data without spectral lines, is used for continuum 
normalization.

The radiative transfer calculations account for atomic hydrogen Rayleigh
and electron scattering.  The mean background radiation field 
is calculated from 24 angles (8 azimuthal with linear weights, 
and 3 zenith per half-sphere with Gaussian weights) 
using a short characteristics scheme, while the emergent flux 
is integrated for 21 angles (vertical ray plus 3 zenithal rays
and 8 azimuthal angles again with linear weights,
except that only 4 zenithal angles are considered 
for the shallowest rays). The abscissae and 
weights for each case can be found 
in Abramowitz \& Stegun (1972). This setup evenly covers a sphere  
with 8 angles for the azimuth (0-360 deg) and 3-4 angles 
for the inclination (0-180 deg). The integration along a 
ray was performed from the top
layer down to optical depths of at least $20$. The final emergent flux
is an average of the flux for all the snapshots considered for each
simulation.

Rotation was not accounted for in this step. It is well-known that rotation
affects the observed line asymmetries (Gray 1986), but low rotational
velocities should only have a modest effect on the net shifts derived by
cross-correlation for the entire Gaia RVS spectral window. In part, this is
related to the spectral resolution of the RVS instrument, 
corresponding to 26 \mbox{km s$^{-1}$} (de Bruijne 2012).

The calculations were performed on the Longhorn cluster at the Texas
Advanced Computing Center (TACC). The intensities for about 150 million
frequencies were derived across all snapshots. The total computational
effort is roughly equivalent to 3 months on a modern workstation with 4
cores.

Fig. \ref{f1} illustrates the spectra in our 1D and 3D grids that
are closest to the atmospheric parameters for Arcturus and the Sun.
This comparison is done after smoothing the data to a FWHM resolving
power of $R = 11,500$ as appropriate for the RVS; 
we adopted  a microturbulent velocity of 1.5 km s$^{-1}$ for the 1D calculations.
The agreement between models and observations is significantly better 
in the solar case
(observed spectrum from Kurucz et al. 1984) than for Arcturus
(Hinkle et al. 2000). Overall, predicted absorption features are 
slightly stronger in 1D than in 3D.
The agreement is reasonable, considering that the models are not 
tailored to these stars. Arcturus has  an $T_{\rm eff}\sim 4300 $ K,
$\log g \simeq 1.7$ (with $g$ in cm s$^{-2}$) and a metallicity about [Fe/H]
$\simeq -0.5$
(see, e.g., Ram\'{\i}rez \& Allende Prieto 2011), and
the Sun has an 
$T_{\rm eff}= 5777 $ K,
$\log g \simeq 4.437$ and, by definition,  
[Fe/H]$=0$ (see, e.g. Stix 2004). The model compared here with
Arcturus  (d3t40g15mm10n01) has an 
$T_{\rm eff}$ of  4040  K,
$\log g = 1.5$ and a metallicity of [Fe/H]$= -1.0$, and 
that compared with the Sun (d3t59g45mm00n01) has an 
$T_{\rm eff}$ of 5865  K,
$\log g = 4.5$ and solar metallicity.

\subsection{1D spectral synthesis}
\label{1ds}

The 1D counterpart radiative transfer calculations were performed
with the 1D version of ASS$\epsilon$T, and therefore the results
can be directly  compared between the 1D and 3D models.
In particular, chemical compositions, opacities, and 
radiative transfer calculation for the emergent intensities
are shared by the 1D and 3D branches. The only  differences between
the 1D and 3D calculations 
are that the opacities are calculated exactly for the grid points
in 1D (every depth in the model atmospheres), while they are interpolated 
in 3D (from a precomputed grid in temperature and density, as described
in \S \ref{3ds}), 
and the radiative transfer
solver for the calculation of the mean intensities in the grid,
which is based on Feautrier's method  (Feautrier 1964) 
in 1D and an integral method in 3D.
Our tests nonetheless indicate that these differences have a negligible
effect on the continuum-normalized fluxes. 
The difference between the Feautrier and the direct method in 1D 
is about 10$^{-5}$. The effect of the interpolation of the opacities 
in 3D is estimated to be on the order of 10$^{-4}$. 
The different choices for the opacity calculation and the radiation transfer method are made to accommodate the very expensive 
calculations in 3D, which are not feasible without interpolating 
the opacities and benefit from the faster 1st order method for 
the integration of the intensities.


\section{Effective convective shifts for the Gaia RVS}

We divided the spectra by the pseudo continua, estimated by
iteratively fitting a low order polynomial to the upper envelope
of each spectrum. We then calculated the net offset between the
3D and 1D spectra to derive the overall convective shift 
predicted by the hydrodynamical simulations by cross-correlation.

The cross-correlation of two  arrays (o spectra)
{\bf T} and {\bf S} of equal and even number of elements $N$ is defined 
as a new array {\bf C} with 
\begin{equation}
 C_i = \sum_{k=1}^{N} T_k S_{k+i-\frac{N}{2}},
\label{Ci}
\end{equation}
\noindent where $i$ runs from $1$ to $N$.
If the spectrum {\bf T} is identical to {\bf S}, but shifted
by an an integer number of pixels $p$, the maximum value in the array {\bf C} 
will correspond to its element $i=p+\frac{N}{2}$. 
Cross-correlation can be similarly
used to measure shifts that correspond to non-integer numbers. 
In this case, finding the location of the maximum value of 
the cross-correlation function
can be performed with a vast choice of algorithms.

We calculated the cross-correlation of 1D and 3D model
spectra using the software by Allende Prieto (2007), which 
allows fitting the peak of the cross-correlation function
with a parabola, a cubic polynomial, or a Gaussian.
Based on tests adding noise to simulated spectra, 
we chose to use parabolic fittings
to the central 3 points around the maximum of 
the cross-correlation function.

\begin{figure}[t!]
\footnotesize
\centering
\includegraphics[width=6cm,angle=90]{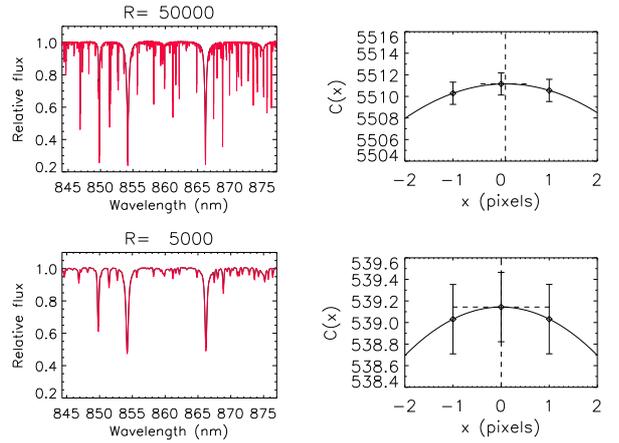}
\caption{The spectra for a single snapshot of model d3t59g45mm00n01 (black) and its
1D counterpart (black), with near solar atmospheric
parameters, are shown in the left-hand panels for two different values of the
resolving power ($R$). The right-hand panels illustrate how the corresponding 
shifts (1D relative to 3D) are obtained
by fitting the peak of the cross-correlation function with a parabola. 
}
\label{xc}%
\end{figure}

\begin{figure}[t!]
\footnotesize
\centering
\includegraphics[width=9cm]{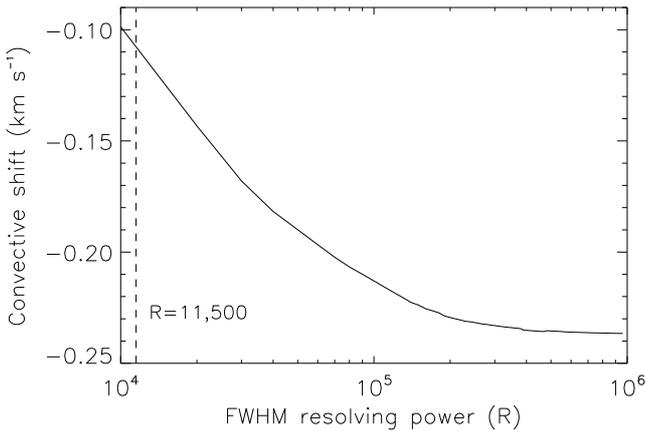}
\caption{Derived convective shift for model d3t59g45mm00n01, with near solar atmospheric
parameters, as a function of the resolving power. These shifts are obtained
by cross-correlation using the Gaia RVS spectral window, and 
fitting the peak of the cross-correlation function with a parabola. The dashed
vertical line marks the RVS resolving power.}
\label{resolution}%
\end{figure}

The net convective shift derived by cross-correlation
 not only depends on the chosen spectral range, but
also on the resolving power of the spectrograph. A degraded
resolution will introduce, in addition to random noise, a
systematic error in the derived velocity shifts. This is illustrated
in Fig. \ref{xc} for a single 3D snapshot and its 1D counterpart at two
different values of the resolving power: resolution does change the
apparent convective shift. Fig. \ref{resolution} shows the derived
shifts (1D relative to 3D) for model d3t59g45mm00n01 (see Table 1 for
reference), which was compared
with the solar spectrum in Fig. \ref{f1}, as a function of the resolving power. 
At the resolving power 
of the Gaia RVS ($R=$11,500), the inferred convective blueshift is about
0.1 km s$^{-1}$, while that measured when the spectral lines
are fully resolved ($R>200,000$) approaches 0.24 km s$^{-1}$.
Our results for the simulations are included in Table 1.
All the values presented in the table, and discussed 
below correspond to the RVS resolution.

In addition to calculating the net convective shifts for the
average spectra from the selected snapshots for each simulation,
we also calculated the shifts for each snapshot and averaged
them out. These correspond to the 'mean' shifts in Table 1.
The variance {\bf ($\sigma^2$)} among snapshots was used to estimate the 
intrinsic uncertainties as $\sigma/\sqrt{n}$, with $n$ the number
of snapshots considered, and these are also included in the table.
We believe that the uncertainties calculated in this way are conservative since the subsample
of snapshots selected for spectral synthesis is chosen to represent the
properties of the complete run (e.g., in effective temperature and mean
velocities). Moreover, the separation in time among the snapshots is large
enough that they can be considered as independent despite being selected from
one time series. 
The agreement between the shifts of the average spectra
and the average shifts from individual snapshots is excellent,
with an average of 0.0004 km s$^{-1}$ and a standard deviation 
of 0.001 km s$^{-1}$.

\begin{figure*}[ht!]
\footnotesize
\centering
\includegraphics[width=16cm]{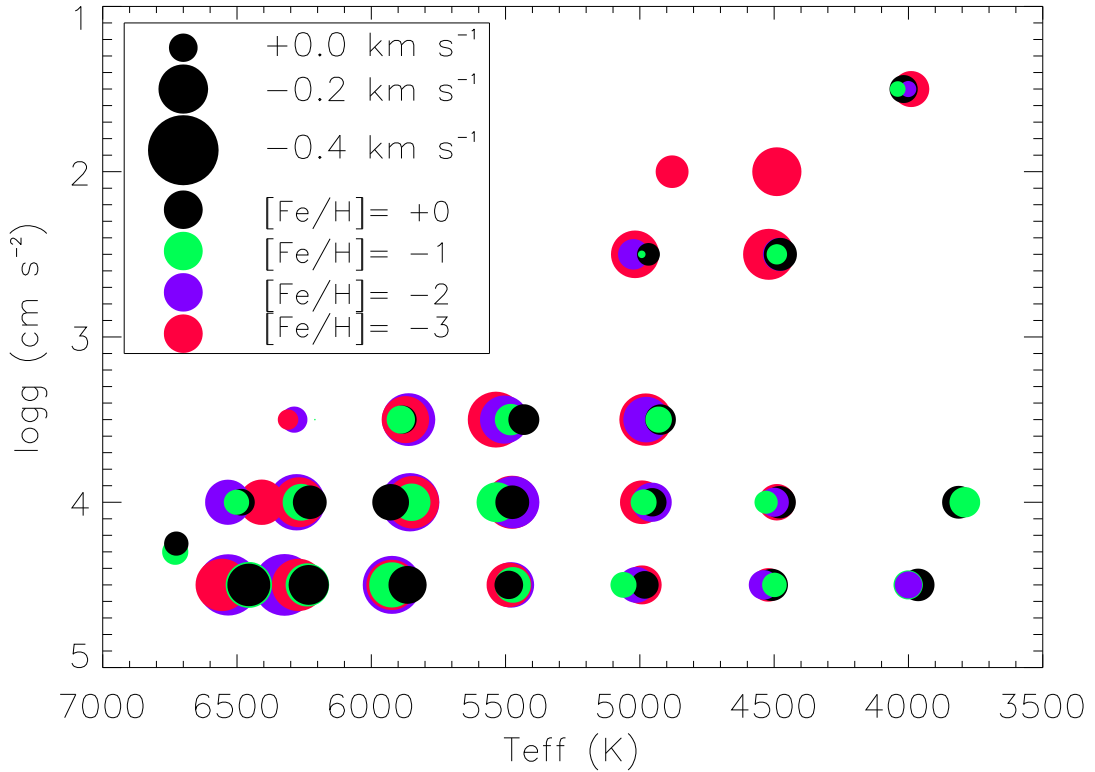}
\caption{Effective convective shifts 
 predicted by the 3D hydrodynamical models for Gaia RVS observations.}
\label{cs}%
\end{figure*}

Fig. \ref{cs} shows the convective shifts for all models.
Note that there are models for four different metallicities, color-coded, 
and that the resulting effective temperatures of the simulations
are not always exactly the same for any combination of
surface gravity and metallicity, as this is not an input  
parameter -- the entropy of the material entering the
simulations box through the open bottom boundary
 is a chosen input instead.

\begin{figure*}[t!]
\footnotesize
\centering
\includegraphics[width=12cm]{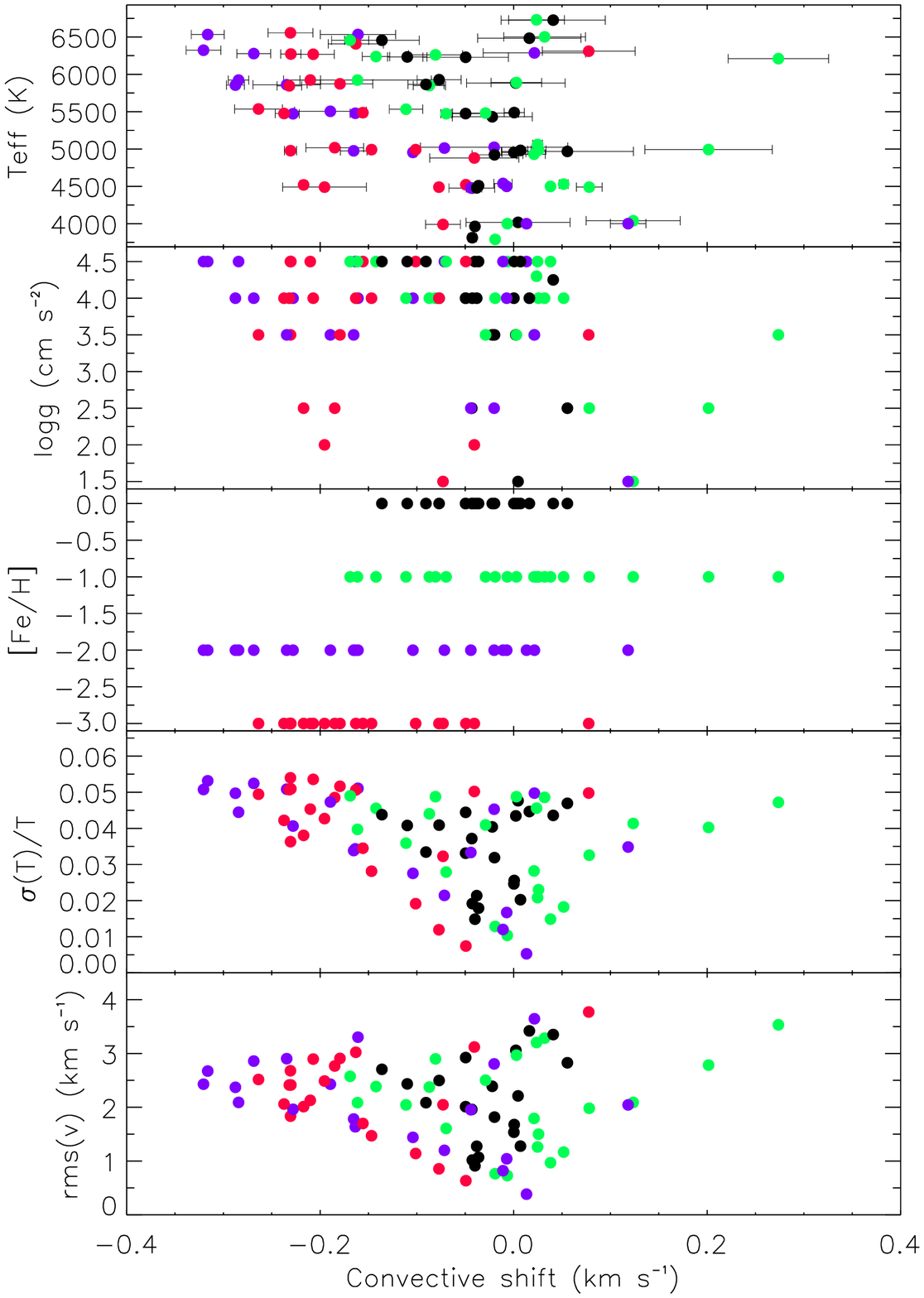}
\caption{Three uppermost panels: Predicted convective shifts and atmospheric
  parameters of the 3D hydrodynamical models for Gaia RVS observations. Two
  lowermost panels: Convective shifts versus relative horizontal temperature
  fluctuations and the rms of the vertical velocity at Rosseland optical depth
  $\tau=2/3$.  In all panels data points are color coded by metallicity as in Fig. \ref{cs}.
}
\label{cs2}%
\end{figure*}

The changes of the net convective shifts with the atmospheric
parameters are perhaps most obvious in Fig.~\ref{cs2}.
The net shifts show a mild correlation with the effective
temperature of the models, in the sense that warmer stars tend to exhibit
larger blueshifts. As found in previous studies 
(Dravins \& Nordlund 1990a,b; Nordlund \& Dravins 1990), 
the net energy flux traversing the atmosphere
is the most important parameter. Surface gravity has a small 
effect on the convective shifts predicted by our simulations
-- overall the shifts intensify at lower gravities --, and so has metallicity, 
with lower net blueshifts at higher metallicities.
Fig. \ref{cs} shows the results obtained when the LHD 1D models,
which are fully consistent with the 3D simulations in terms of
opacities and the equation of state, are used as reference.
However, very similar results are found when  Kurucz models are
used as a reference (see below).

We adopted a micro-turbulence of 2.0 km s$^{-1}$, and a macro-turbulence of
1.5 km s$^{-1}$ for the reference 1D calculations, but our experiments
indicate that these parameters would only lead to a modest impact on the spectral
shifts determined by cross-correlation. For instance, completely neglecting
macro-turbulence leads to a mean correction in the predicted convective
blueshifts of just 0.01 ($\sigma=0.032$) km s$^{-1}$.  Adopting a
micro-turbulence of 1.5 km s$^{-1}$ instead of 2.0 km s$^{-1}$ will, on
average, reduce the convective blueshifts by about 0.007 ($\sigma= 0.011$) km
s$^{-1}$. Finally, adopting Kurucz model atmospheres instead of LHD models as
our 1D reference will, on average, enhance the blueshifts by about 0.01
($\sigma=0.03$) km s$^{-1}$.

The smallest (in absolute terms) convective shifts in the grid are for the
coolest dwarfs of low metal content, and amount barely to $<0.1$ km s$^{-1}$.
The most vigorous net velocities are found for low metallicity F-type
subgiants, practically the warmest stars considered in the grid, where they
exceed 0.3 km s$^{-1}$. 

There are a few simulations (mostly with [Fe/H]$=-1$)
for which redshifts are predicted.  
While perhaps not very evident in Fig.~\ref{cs2} the drop is
clearly apparent in a global analytical fit of the convective shifts of the 3D simulations 
(see Appendix~\ref{sec:fittingfunction}). 
These redshifts are  a clear prediction of the simulations. 
They are caused by strong redshifts of the cores of the Ca\,II triplet lines, 
but they dominate the overall cross-correlation signal.
The cores get redshifted when the lines are strong and the flow exhibits a 
pronounced "reverse granulation" pattern. Still, the overall shift of 
the Ca\,II triplet line cores is a superposition of redshifted and 
blueshifted contributions and therefore it is difficult to predict a 
priori which ones are dominating.

We tried to interpret our findings in terms of the temperature and velocity
fluctuations present in the 3D simulations. We averaged the temperature and
vertical velocity component horizontally and temporally on surfaces of
constant Rosseland optical depth. We took values at Rosseland depth $\tau=2/3$
as representative for the conditions prevailing in the line forming regions --
at least for weak lines. These are shown in the two bottom panels of Fig. \ref{cs2}, 
where a  mild correlation between temperature and velocity fluctuations
is apparent.

\begin{figure}
\footnotesize
\centering
\includegraphics[width=8.0cm]{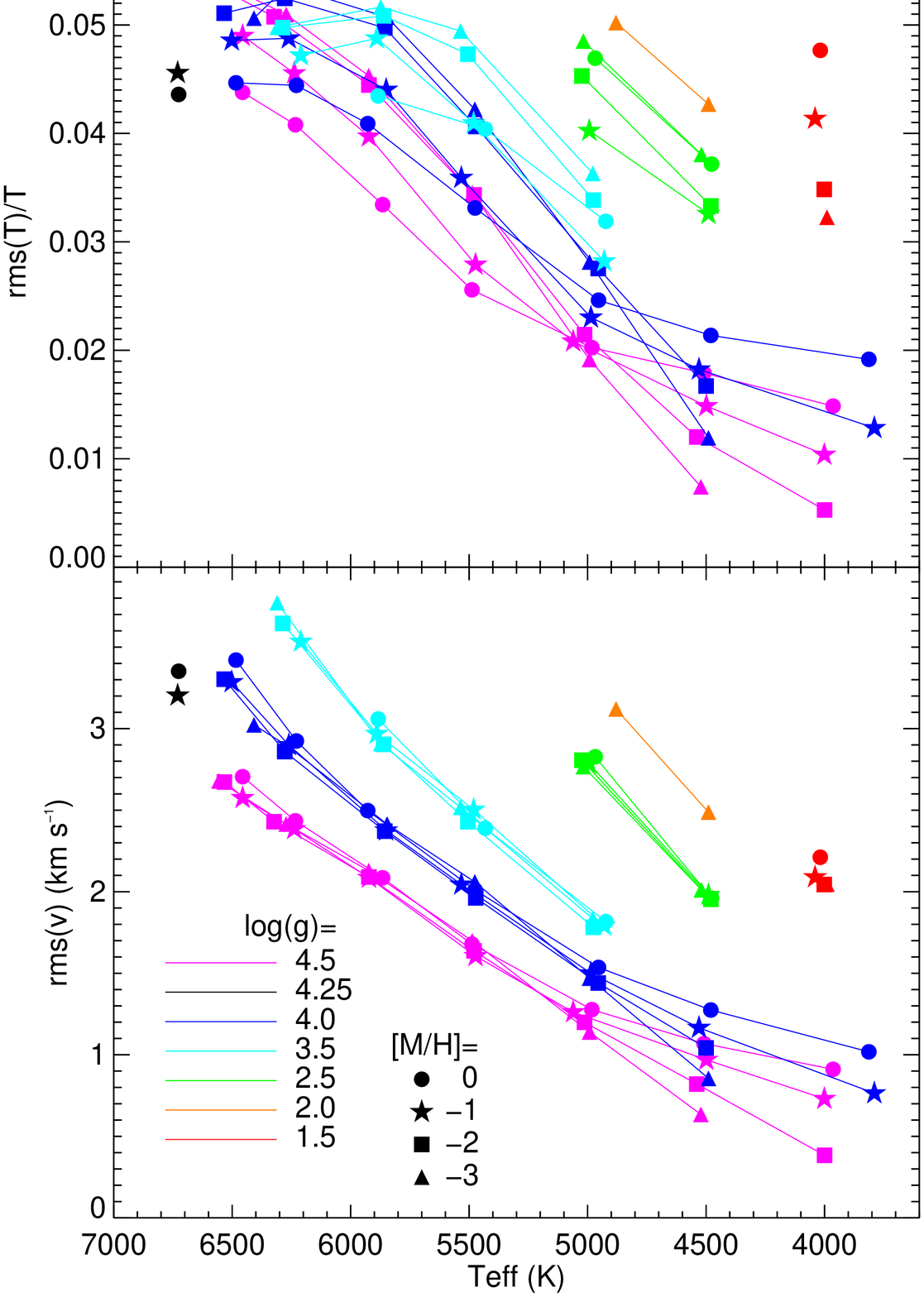}
\caption{Relative horizontal temperature fluctuation (top panel) and absolute
  vertical velocity fluctuations (bottom panel) at Rosseland optical depth
  $\tau=2/3$ in the 3D models. The surface gravity is color coded, metallicity
  is depicted by different plot symbols.
 }
\label{avgs2}%
\end{figure}

Figure~\ref{avgs2} shows how the temperature and velocity fluctuations
depend on effective temperature, surface gravity and metallicity. 
Interestingly, we find that the velocity fluctuations depend on metallicity 
for $\Teff<5000\pun{K}$ but they do not for warmer stars.
Only close to the main-sequence a
systematic dependence appears below 5000\pun{K} with higher metallicity
implying greater fluctuations. For main-sequence stars, $\Teff=5000\pun{K}$
also separates regions of different correlation between metallicity and
temperature fluctuations. 
Above 5000 K, higher metallicity corresponds to
lower temperature fluctuations but this is reversed below 5000 K. While
Fig.~\ref{avgs2} certainly motivates the increase in convective shifts
towards higher effective temperatures and lower gravity, it cannot explain
all trends of the convective shifts. What is indeed missing is information
on the mutual correlation between temperature and velocity fluctuations, as
well as on changes of the line formation conditions with atmospheric parameters
-- as discussed further below. In particular, Fig.~\ref{avgs2} does not
explain the fast drop of the predicted convective shifts near the highest 
effective temperatures in our grid.

Nevertheless, our results resemble earlier high-resolution studies of spectral
line shapes and convective shifts, but the net velocity shifts predicted for
the RVS observations are smaller than for individual, resolved, spectral
lines  (see, e.g., 
Bigot \& Thevenin 2006, 2008; Chiavassa et al. 2011).  
As expected, the net shift measured by cross-correlation is a weighted
average of the lineshifts from all spectral features.  At high metallicity,
many lines contribute, with strong lines giving very small convective shifts,
and weaker features contributing larger shifts. For example, for a solar-like
atmosphere, if we apply cross-correlation by isolating windows around the Ca
II lines, we find very small convective shifts on the order of 0.030 km
s$^{-1}$, while limiting the spectral window to the stretch between the Ca II
lines (857-863 nm), the net shift is about $-0.18$ km s$^{-1}$, double than
what we obtain for the whole RVS spectral range.

For a solar 3D model not included in the grid discussed here, we found in a
previous paper (Allende Prieto et al. 2009) a convective shift of $-0.26$ km
s$^{-1}$ , in excellent agreement with the value directly derived from the
solar flux atlas of Kurucz et al. (1984). These two values correspond to
observations at much higher spectral resolution, and can be compared with
those we find (see Fig. 2) for the model in the grid with parameters closest
to solar, d3t59g45mm00n01, namely $\simeq -0.24$ km s$^{-1}$ when the
resolving power is $R>200,000$ but note that changes to $-0.09$ km s$^{-1}$ at $R=11,500$.

From a theoretical standpoint, the main effect of a reduced gravity is a lower
surface pressure, as well as an increase in both the vertical and horizontal
scales. At a given effective temperature the flow is constrained 
to transport the
same energy flux so that a reduced pressure and density (since the temperature
is almost fixed) has to be compensated for by an increase in the velocity
and/or the temperature fluctuations (Dravins \& Nordlund 1990b; Dravins et
al. 1993; Collet, Asplund \& Trampedach 2007; Gray 2009). As a result,
convective blueshifts should strenghten -- as long as the surface area
fractions covered by up-flows and down-flows do not change. As illustrated
in Fig. \ref{avgs2},  our
detailed simulations roughly support this simple picture.  

\begin{figure}[h!]
\footnotesize
\centering
\includegraphics[width=9cm]{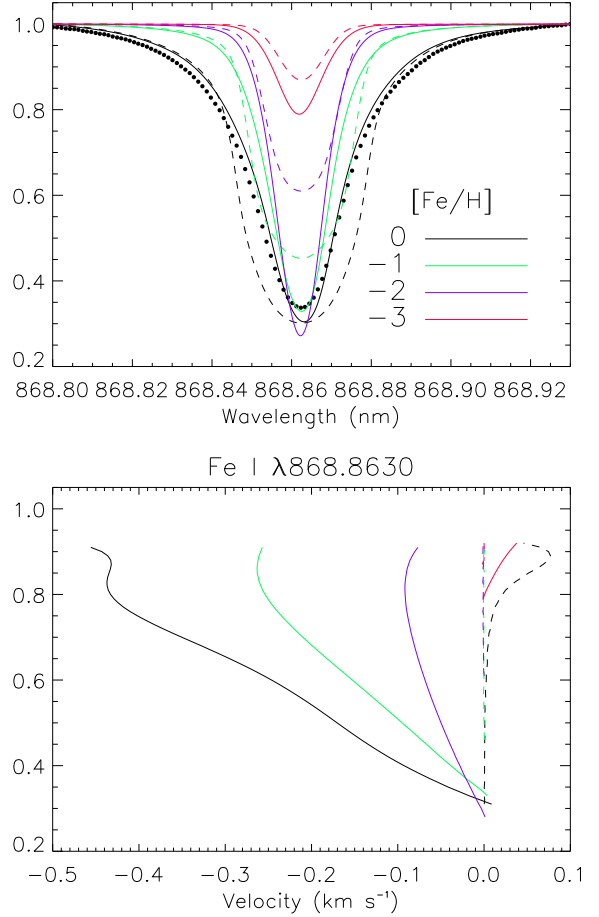}
\caption{Line profiles and velocity bisectors for the Fe I transition
at 868.6 nm. The filled circles in the upper panel correspond to solar observations.
 Solid lines are for 3D simulations and broken lines for 1D models. Note that 
 the 1D calculation at solar-metallicity for this line (broken solid line) 
 shows a non-zero velocity bisector, due to overlapping CN transitions included
 in all our calculations that distort the otherwise perfectly symmetric 
 1D profiles.}
\label{bisector}%
\end{figure}

At lower metallicity we expect the reduced continuum opacity to make visible
deeper atmospheric layers with more vigorous convection (Allende Prieto et
al. 1999), and the simulations indicate that the convective shifts of
individual lines become stronger, but the span of the line bisectors weakens
dramatically with metallicity.
This is illustrated for the strong Fe I line at 868.86 nm in solar-like
($T_{\rm eff}\simeq 5900$ K and $\log g =4.5$) stars in Fig. 
\ref{bisector}. In this figure, we show line profiles (top panel) and
the corresponding bisectors (lower panel), with the solid lines  for  
the calculations based on 3D models and the dashed lines for
1D models. This analysis is performed at full resolution,
since line bisectors are erased at the  RVS resolution, using
a micro-turbulence of 2 km s$^{-1}$ and no macro-turbulence.
As the feature weakens, the range of atmospheric heights
sampled from the wings to the core is reduced, which is a likely explanation for 
the smaller bisector spans. For this line, we also find that the predicted 
convective shift is $-0.00$, $-0.10$, $-0.12$, and $-0.13$ km s$^{-1}$ at 
[Fe/H]$=0, -1, -2$, and $-3$, respectively. Note that these bisectors are
calculated using the standard definition, i.e. subtracting the velocity
in the blue wing from that at the same normalized flux level in the 
red wing, and thus the net line shifts are not included and they
approach zero velocity in the line core (Gray 1992).
In addition to the characteristics
of surface convection changing as a function of metallicity, 
the strengthening of the line blueshifts
with decreasing metallicity is  related to the line formation
shifting deeper into the photosphere as the line weakens. This is
consistent with our findings for the overall net blueshift for 
Gaia RVS observations.

The filled circles in the upper panel of Fig. \ref{bisector} show the renormalized
observations for the Sun of Kurucz et al. (1984). A significant improvement 
in the agreement between theory and observation is obvious for the line core 
when going from 1D to 3D models (dashed and solid black lines, respectively).
A similar effect is found using a Kurucz model, 
instead of its LHD counterpart, 
for the 1D spectrum. It is also noticeable the extreme discrepancies between
the 1D and 3D cases at [Fe/H]$=-2$, as it is the large
reduction in strength for the line calculated in 3D 
between [Fe/H]$=-2$ and $-3$. An exhaustive check of these
predictions against high quality observations of metal-poor stars
is very much needed.

Our calculations are likely least reliable for cool giants, and in fact the
predicted convective shifts for stars approaching 4000 K are unlikely to be
realistic.  Gray, Carney \& Yong (2008) examined bisectors for a number of red
giant stars with low metallicity, finding that those with temperatures of
about 4100 K or less tend to exhibit reversed bisectors (with an 'inverse C'
shape, rather than with a 'C'-like shape seen in solar-type stars). More
recent results for the supergiant $\gamma$\,Cyg confirm the same trend (Gray
2010).  These stars also tend to exhibit radial velocity 'jitter' (see, e.g.,
Carney et al. 2008, and the theoretical predictions by Chiavassa et al. 2011).

We have 
identified an expression that can reproduce the net convective blueshifts
theoretically predicted for the RVS observations in Table 1 with an rms of
0.021 km s$^{-1}$; and a maximum deviation of 0.074 km s$^{-1}$.  
We have coded this expression into an IDL function
provided in the appendix.

\section{Gravitational redshifts}

Photons escaping from a stellar photosphere to infinity are redshifted 
as a result of climbing up the gravitational field by 
$Gc M/R$, where $G$ is the Gravitational constant, $c$ the speed of light,
 $M$ the stellar mass, and $R$ the stellar radius. Using solar units for 
the stellar mass and radius, the gravitational redshift amounts to
0.6365 $M/R$ km s$^{-1}$, and for light intercepted at 1 AU
0.6335 $M/R$ km s$^{-1}$ (Lindegren \& Dravins 2003). 
The same  effect acts when photons
enter in the Earth's gravitational field, which reduces the redshift
by about 0.2 m s$^{-1}$, but this small correction can be ignored for most purposes.
With the parallaxes and spectrophotometry
provided by Gaia, it should be possible to obtain fairly accurate 
customized predictions for the gravitational redshifts for individual Gaia 
targets. 

Gravitational shifts for late-type dwarfs amount to about 0.7-0.8 km s$^{-1}$
at $\log g \sim 4.5$,  but they dramatically decrease with surface gravity down 
to 0.02--0.03 km s$^{-1}$ for the giants in our grid ($\log g \sim 1.5$).

\section{Discussion and Summary}

The radial velocities inferred from the wavelength shifts
measured in stellar spectra are systematically
offset from the true space velocities of stars projected 
along the line of sight
due to a number of effects. Quantifying the impact of such
effects is model dependent, and therefore the IAU has introduced
the term 'radial-velocity measure' (Lindegren \& Dravins
2003) to refer to the radial velocity naively inferred
from the relative wavelength shifts of lines
$c \delta \lambda/\lambda$, 
and encourages observers to keep  measurements disentangled
from subsequent corrections\footnote{We note that the 'radial-velocity measure' does not have a unique definition, since the value derived depends
on the spectral features and the actual method used. Therefore, such measures
must be accompanied by a detailed description of the procedure used to
derive it.}.
Nevertheless, for most stars, two relatively clean corrections
are dominant: photospheric gravitational redshifts and
convective blueshifts. In the case of stars moving at high
speed in the direction perpendicular to the observer, the relativistic
contribution to the Dopper shift needs to
 be considered at a precision level of hundred of meters
per second; the spectrum of a star with a transverse 
velocity of 300  km s$^{-1}$ and no radial velocity 
will be redshifted as much as that of a star with a receding 
radial velocity of 0.15 km s$^{-1}$.

We have used three-dimensional radiative-hydrodynamical simulations of
surface convection in late-type stars to evaluate the impact of
convective  shifts on the spectra from the Radial Velocity Spectrometer
onboard Gaia. The net velocity shifts derived by cross-correlation 
depend both on the wavelength range and the spectral resolution of the
observations. Our models predict very small net convective shifts
for the RVS observations of K-type stars, which increase with effective
temperature to reach bout 0.3 km s$^{-1}$ for F-type dwarfs. The
predicted shifts depend as well on metallicity and surface gravity. 

In the case of the Gaia mission, the corrections discussed above, which are
$<1$ km s$^{-1}$ are  small in comparison with the typical 
uncertainties in the space velocities determined for individual stars, 
but can cause systematic errors when average values are derived 
for ensembles of stars. Even more important, since the Gaia RVS
is a self-calibrated instrument, in the sense that the wavelength solution is
derived from stellar observations, systematic offsets from star to
star such as those related to convective shifts, need to be accounted
for in the data reduction. In this paper, we used
a grid of 3D hydrodynamical simulations to  estimate 
convective blueshifts for stars with spectral types
F to M, at different metallicities and evolutionary stages.  
We note that from data provided
by Gaia itself, it will be possible to constrain the stellar masses and
radii for individual stars, and therefore the gravitational
redshifts.

Gravitational redshifts dominate convective
blueshifts predicted for the Gaia window for dwarfs of all temperatures, 
in particular for later spectral types (K and M), where the convective 
shifts are weakest. Gravitational redshifts become weaker
for stars with lower gravities. The opposite tendency, if less 
pronounced, is expected for the convective blueshifts, in such a way
that gravitational and convective shifts nearly cancel each other
for stars with intermediate gravities. 

When adding together convective and gravitational shifts, we
expect the largest velocity offset for the coolest dwarfs
in our study, with effective temperatures around 4000 K, 
where the net effect, dominated by gravitational redshifts,
amounts up to nearly 0.6 km s$^{-1}$.
By using the corrections provided here and custom-made
estimates of the gravitational shifts, these sources
of systematic error should be reduced significantly to a level 
ranging from about $0.1$ km s$^{-1}$ for F-type dwarfs, to 
a few tens of meters per second for late-K-type stars. 

In addition to calculated convective shifts, we provide
FITS files including the 1D and 3D model fluxes we 
have computed. These may turn
useful to build radial velocity templates for derivation
of radial velocities from observations.

We end with a word of caution. The theoretical calculations presented 
here only have been checked against a limited set of observations, 
mainly for the Sun. An exhaustive check against high-quality observed
spectra for warmer and cooler temperatures, especially at very low
metallicity, is still needed before one can ensure that the predicted
shifts are completely trustworthy.

\begin{acknowledgements}
	We are indebted to Ivan Hubeny for assistance computing radiative opacities.
      Carlos Allende Prieto is thankful to 
      Mark Cropper, David Katz, and Fr\'ed\'eric Th\'evenin for
      fruitful discussions.
      B.F. acknowledges financial support from
		the {\sl Agence Nationale de la Recherche} (ANR), and the
	  {\sl ``Programme National de Physique Stellaire''} (PNPS) of
          CNRS/INSU, France. 
HGL acknowledges financial support by the Sonderforschungsbereich SFB\,881
``The Milky Way System'' (subproject A4) of the German Research Foundation
(DFG).
\end{acknowledgements}

\appendix
\section{An empirical fitting function to describe the blueshifts as a function of atmospheric
parmeters}
\label{sec:fittingfunction}

To derive from the tabulated data an easily applicable function
we follow the recursive fitting procedure outlined by Sbordone et al. (2010):
we searched for an analytical function for the convective blueshift
with three independent parameters ($\log T_{\rm eff}$,
$\log g$ and [Fe/H]), written as   $y = y(x_0, x_1, x_2; \vec{A})$.
The function and the vector of free parameters $\vec{A}=(A_0, A_1,\ldots)$
that give the best fit to the tabulated data is to be found.

The function-find routine (``fufi'', written in IDL) starts with
the most simple ``function'', a constant $A_0$ --
the weighted average of the tabulated blueshift data.
At each recursive iteration we let the routine replace each parameter $A_i$
of a candidate fitting function  with a polynomial
  $A_i \rightarrow A_0 + A_1 x_0 + A_2 x_1 + A_3 x_2$
or, alternatively, an exponential
  $A_i \rightarrow A_0 + A_1 \exp(A_2 + A_3 x_0 + A_4 x_1 + A_5 x_2)$
to reach the next level of complexity.
For each of these functions, the optimum parameter set is then searched, using the
parameters from the previous level as starting point.
We use the inverse of the rms fluctuations 
of the convective shifts between the snapshots for a given stellar parameter combination (or 0.01 km s${-1}$ when the fluctuations are below this value)
as fitting weights. The function with the fitting-parameter set that provides the smallest overall error (among thousands of candidates)
is finally written to a file as a Fortran or IDL function.
The function we supply has been slightly edited to improve readability.
We emphasize that the functional form and fitting parameters have no
physical meaning whatsoever.

\begin{figure}
\footnotesize
\centering
\includegraphics[width=8.8cm]{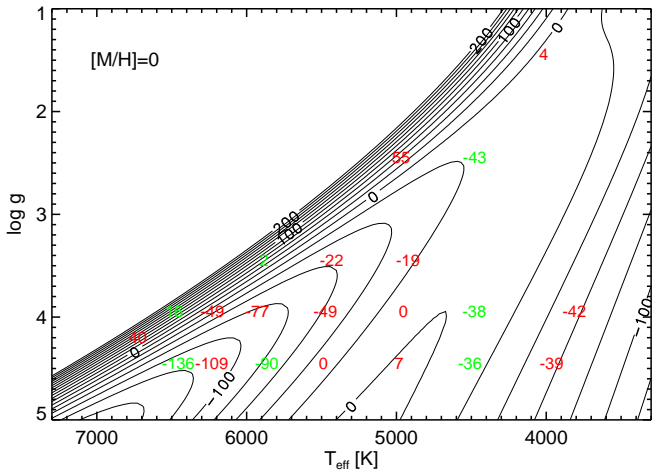}
\caption{Convective shifts for solar metallicity simulations (in units of
m s$^{-1}$). The contours join
the locations with the same convective blueshift, as predicted
by our analytical fitting function. Red/green color is used in the
regions where the analytical function under/over predicts the measurements 
performed on the simulations.}
\label{csfit0}
\end{figure}

\begin{figure}
\footnotesize
\centering
\includegraphics[width=8.8cm]{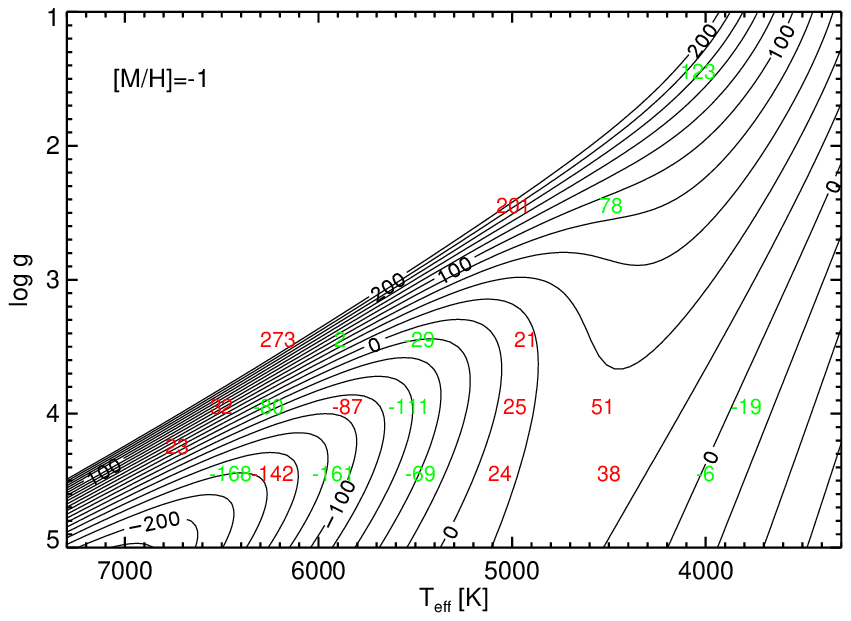}
\caption{Convective shifts, \moh=-1. Similar to Fig. \ref{csfit0}.}
\label{csfit1}
\end{figure}

\begin{figure}
\footnotesize
\centering
\includegraphics[width=8.8cm]{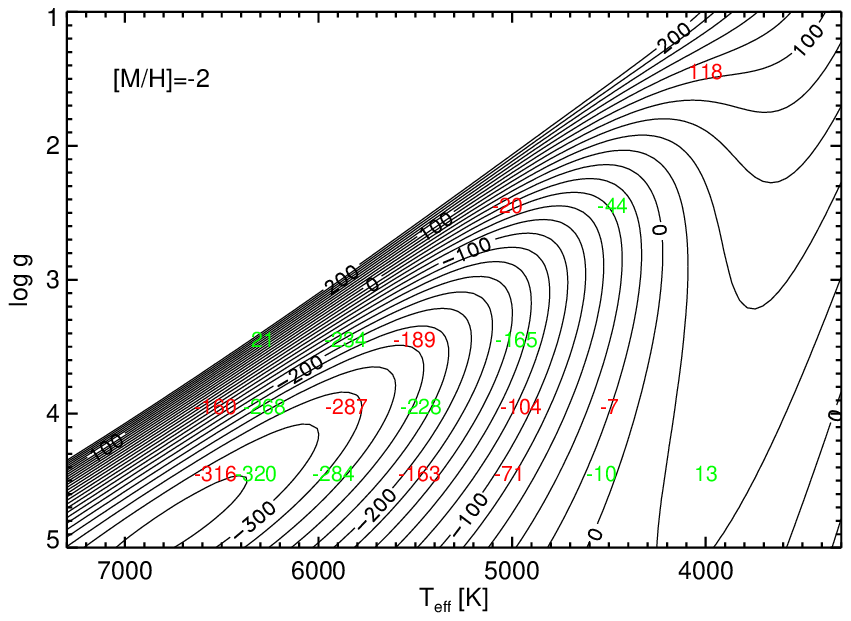}
\caption{Convective shifts, \moh=-2. Similar to Fig. \ref{csfit0}.}
\label{csfit2}
\end{figure}

\begin{figure}
\footnotesize
\centering
\includegraphics[width=8.8cm]{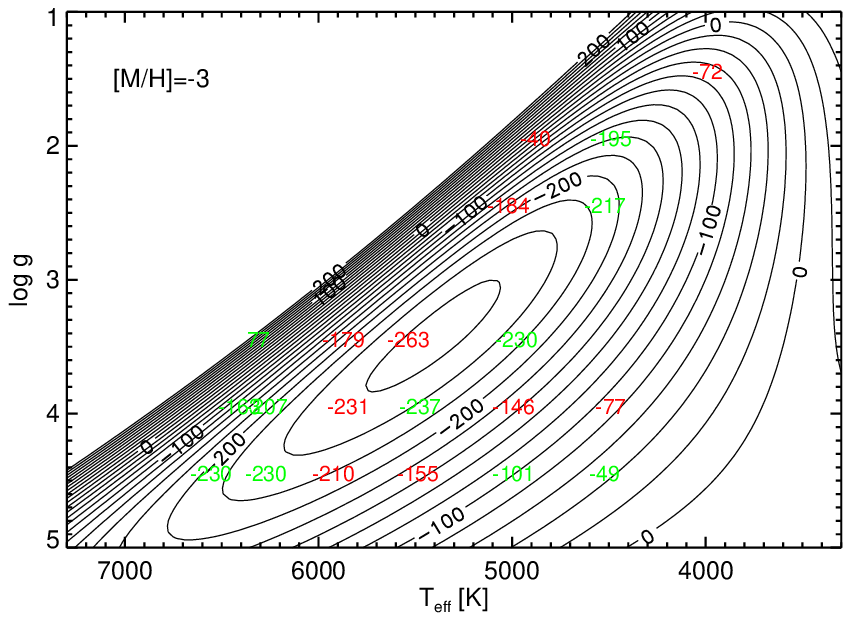}
\caption{Convective shifts, \moh=-3.  Similar to Fig. \ref{csfit0}.}
\label{csfit3}
\end{figure}

We varied some control parameters (the list of candidate terms and the
recursion depth), and compared errors and the overall functional dependence of
the results.  Fitting functions composed exclusively of simple polynomial
terms already give decent fits (and would require a much simpler procedure to
derive than the one outlined above).  Still, fits including the exponential
term are superior and behave much better and closer to our expectations at --
and even slightly beyond -- the borders of the set of stellar parameters for
which model results are available. The overall rms scatter 
between the direct measurements from the simulations and the 
fitting formula is just 0.021 km s$^{-1}$, with a maximum deviation of 0.074 km s$^{-1}$. 
We give an implementation of the fitting
function in IDL below. It includes a check that the input parameters are
within the range covered by the 3D models. A few of the free fitting
parameters are zero, but are left in the routine to allow for a more systematic
writing of the final expression. Figs. \ref{csfit0}, \ref{csfit1}, \ref{csfit2}, 
\ref{csfit3} illustrate the convective shifts predicted by the
fitting equation for [Fe/H]$=0, -1, -2$, and $-3$, respectively. 

\vspace{1cm}

{\tt

function ConvShift, Teff, logg, FeH

;

;	IN: Teff -- effective temperature (K)

;		logg -- surface gravity (g in cm/s2)

;		FeH  -- [Fe/H] metallicity relative to solar 

;

;	OUT: Convective shift (m/s) 

;   (<0 for blueshifts)

;

; --- Thu Oct 20 20:09:50 2011 --- B.F. ---

;check that 3 parameters are input

if n\_params() lt 3 then begin

	print,'

	return,-10000000000.d0

endif

;check that we're within limits

if (min(Teff) lt 3790.00 or \$

max(Teff) gt 6730.00 or \$

min(FeH) lt -3.0 or \$

max(FeH) gt 0. or \$

max(logg) gt 4.5 or \$

min(logg-(teff*9.184e-4-2.482)) lt 0.0) \$

then begin

	~~~print,'\% ConvShift: Params. out of range:'

	~~~print,'\%   3790 <= Teff <= 6730. K'

	~~~print,'\%   9.184e-4*Teff-2.482<=logg<= 4.5'

	~~~print,'\%  -3.0 <= [Fe/H] <= 0.0 dex'

	~~~return,-10000000000.d0

endif

logTeff=alog10(Teff*1d-3)

M=-FeH

A=dindgen(24)

A(00)= 3.9921954E-01 \&
A(01)=-7.1438992E-01

A(02)= 2.0028012E-02 \&
A(03)=-5.2146580E-02

A(04)=-4.1100047E-06 \&
A(05)= 0.0000000E+00
 
A(06)= 2.2334658E+01 \&
A(07)=-1.7070215E+00

A(08)=-1.7445524E-01 \&
A(09)= 1.4332379E-02

A(10)= 0.0000000E+00 \&
A(11)=-1.3239030E+01

A(12)= 8.2077384E-01 \&
A(13)=-9.6311294E-02

A(14)=-1.7150490E+03 \&
A(15)= 0.0000000E+00

A(16)=-9.3405848E+00 \&
A(17)= 7.7603954E-01

A(18)= 1.3519228E+00 \&
A(19)=-5.0111694E+00

A(20)= 0.0000000E+00 \&
A(21)=-8.2077414E-01

A(22)= 1.0457402E-01 \&
A(23)= 4.2118204E-01

Shift=A(00)+A(01)*logTeff+(A(02)  + \$

(A(04)+A(09)*exp(A(10)+(A(11)+ \$

A(14)*exp(A(15)+ \$

(A(16)+A(19)*exp(A(20)+ \$

A(21)*logTeff+ \$

A(22)*logg+ A(23)*M))*logTeff+ \$

A(17)*logg+A(18)*M))*logTeff+ \$

A(12)*logg+A(13)*M))*exp(A(05)+ \$

A(06)*logTeff+A(07)*logg+A(08)*M))*logg+A(03)*M

Shift=-Shift

return, Shift

end 
}

\clearpage

\onecolumn
\begin{longtable}{lllrrrr}

  \caption[]{Predicted  convective line shifts. The fifth column gives the
  convective shifts measured in the average spectrum from all snapshots considered,
   and the sixth column gives the average of the shifts measured in individual snapshots.}
         \label{t1}
\\
 {\rm Simulation ~ID }   & T$_{\rm{eff}}$ &  $\log g$  & {\rm [Fe/H]}  & 
 {\rm Convective~ shift}
 & {\rm Mean} & {\rm Error} \\
 & (K)         &  (cm s$^{-2}$) & (dex) & (km s$^{-1}$) 
                    & (km s$^{-1}$) & (km s$^{-1}$)  \\
            \hline
d3t38g40mm00n2 & 3813 &  4.00 & $ -0.00$ & $ -0.043$ & $ -0.043$ & $  0.004$ \\
d3t38g40mm10n3 & 3790 &  4.00 & $ -1.00$ & $ -0.019$ & $ -0.019$ & $  0.001$ \\
d3t40g15mm00n02 & 4018 &  1.50 & $ -0.00$ & $  0.000$ & $  0.005$ & $  0.054$ \\
d3t40g15mm10n01 & 4040 &  1.50 & $ -1.00$ & $  0.123$ & $  0.124$ & $  0.049$ \\
d3t40g15mm20n01 & 4001 &  1.50 & $ -2.00$ & $  0.119$ & $  0.118$ & $  0.018$ \\
d3t40g15mm30n01 & 3990 &  1.50 & $ -3.00$ & $ -0.073$ & $ -0.073$ & $  0.018$ \\
d3t40g45mm00n01 & 3964 &  4.50 & $ -0.00$ & $ -0.040$ & $ -0.040$ & $  0.002$ \\
d3t40g45mm10n01 & 4001 &  4.50 & $ -1.00$ & $ -0.007$ & $ -0.007$ & $  0.001$ \\
d3t40g45mm20n01 & 4000 &  4.50 & $ -2.00$ & $  0.013$ & $  0.013$ & $  0.000$ \\
d3t45g20mm30n01 & 4490 &  2.00 & $ -3.00$ & $ -0.196$ & $ -0.195$ & $  0.043$ \\
d3t45g25mm00n01 & 4477 &  2.50 & $ -0.00$ & $ -0.037$ & $ -0.043$ & $  0.024$ \\
d3t45g25mm10n01 & 4490 &  2.50 & $ -1.00$ & $  0.078$ & $  0.078$ & $  0.013$ \\
d3t45g25mm20n02 & 4480 &  2.50 & $ -2.00$ & $ -0.044$ & $ -0.044$ & $  0.005$ \\
d3t45g25mm30n02 & 4520 &  2.50 & $ -3.00$ & $ -0.218$ & $ -0.217$ & $  0.004$ \\
d3t45g40mm00n01 & 4480 &  4.00 & $ -0.00$ & $ -0.038$ & $ -0.038$ & $  0.005$ \\
d3t45g40mm10n01 & 4530 &  4.00 & $ -1.00$ & $  0.052$ & $  0.052$ & $  0.005$ \\
d3t45g40mm20n01 & 4500 &  4.00 & $ -2.00$ & $ -0.007$ & $ -0.007$ & $  0.001$ \\
d3t45g40mm30n02 & 4490 &  4.00 & $ -3.00$ & $ -0.077$ & $ -0.077$ & $  0.001$ \\
d3t45g45mm00n01 & 4509 &  4.50 & $ -0.00$ & $ -0.036$ & $ -0.036$ & $  0.004$ \\
d3t45g45mm10n01 & 4499 &  4.50 & $ -1.00$ & $  0.038$ & $  0.038$ & $  0.002$ \\
d3t45g45mm20n01 & 4539 &  4.50 & $ -2.00$ & $ -0.011$ & $ -0.011$ & $  0.009$ \\
d3t45g45mm30n01 & 4522 &  4.50 & $ -3.00$ & $ -0.050$ & $ -0.049$ & $  0.000$ \\
d3t49g20mm30n01 & 4880 &  2.00 & $ -3.00$ & $ -0.042$ & $ -0.041$ & $  0.046$ \\
d3t50g25mm00n01 & 4968 &  2.50 & $ -0.00$ & $  0.054$ & $  0.056$ & $  0.068$ \\
d3t50g25mm10n01 & 4993 &  2.50 & $ -1.00$ & $  0.198$ & $  0.201$ & $  0.066$ \\
d3t50g25mm20n01 & 5024 &  2.50 & $ -2.00$ & $ -0.020$ & $ -0.020$ & $  0.076$ \\
d3t50g25mm30n01 & 5018 &  2.50 & $ -3.00$ & $ -0.187$ & $ -0.185$ & $  0.030$ \\
d3t50g35mm00n01 & 4923 &  3.50 & $ -0.00$ & $ -0.019$ & $ -0.020$ & $  0.023$ \\
d3t50g35mm10n01 & 4930 &  3.50 & $ -1.00$ & $  0.021$ & $  0.021$ & $  0.012$ \\
d3t50g35mm20n01 & 4976 &  3.50 & $ -2.00$ & $ -0.165$ & $ -0.165$ & $  0.014$ \\
d3t50g35mm30n01 & 4978 &  3.50 & $ -3.00$ & $ -0.231$ & $ -0.231$ & $  0.006$ \\
d3t50g40mm00n01 & 4954 &  4.00 & $ -0.00$ & $  0.000$ & $  0.000$ & $  0.012$ \\
d3t50g40mm10n01 & 4986 &  4.00 & $ -1.00$ & $  0.026$ & $  0.026$ & $  0.008$ \\
d3t50g40mm20n01 & 4955 &  4.00 & $ -2.00$ & $ -0.104$ & $ -0.104$ & $  0.004$ \\
d3t50g40mm30n01 & 4992 &  4.00 & $ -3.00$ & $ -0.147$ & $ -0.147$ & $  0.003$ \\
d3t50g45mm00n04 & 4982 &  4.50 & $ -0.00$ & $  0.007$ & $  0.007$ & $  0.008$ \\
d3t50g45mm10n03 & 5061 &  4.50 & $ -1.00$ & $  0.025$ & $  0.025$ & $  0.005$ \\
d3t50g45mm20n03 & 5013 &  4.50 & $ -2.00$ & $ -0.072$ & $ -0.072$ & $  0.002$ \\
d3t50g45mm30n03 & 4992 &  4.50 & $ -3.00$ & $ -0.101$ & $ -0.101$ & $  0.001$ \\
d3t55g35mm00n01 & 5432 &  3.50 & $ -0.00$ & $ -0.022$ & $ -0.022$ & $  0.041$ \\
d3t55g35mm10n01 & 5481 &  3.50 & $ -1.00$ & $ -0.030$ & $ -0.029$ & $  0.034$ \\
d3t55g35mm20n01 & 5505 &  3.50 & $ -2.00$ & $ -0.191$ & $ -0.189$ & $  0.037$ \\
d3t55g35mm30n01 & 5536 &  3.50 & $ -3.00$ & $ -0.265$ & $ -0.264$ & $  0.025$ \\
d3t55g40mm00n01 & 5475 &  4.00 & $ -0.00$ & $ -0.050$ & $ -0.050$ & $  0.018$ \\
d3t55g40mm10n01 & 5533 &  4.00 & $ -1.00$ & $ -0.111$ & $ -0.111$ & $  0.017$ \\
d3t55g40mm20n01 & 5472 &  4.00 & $ -2.00$ & $ -0.228$ & $ -0.228$ & $  0.006$ \\
d3t55g40mm30n01 & 5476 &  4.00 & $ -3.00$ & $ -0.237$ & $ -0.237$ & $  0.009$ \\
d3t55g45mm00n01 & 5488 &  4.50 & $ -0.00$ & $  0.000$ & $  0.001$ & $  0.010$ \\
d3t55g45mm10n01 & 5473 &  4.50 & $ -1.00$ & $ -0.070$ & $ -0.070$ & $  0.005$ \\
d3t55g45mm20n01 & 5479 &  4.50 & $ -2.00$ & $ -0.164$ & $ -0.164$ & $  0.005$ \\
d3t55g45mm30n01 & 5487 &  4.50 & $ -3.00$ & $ -0.156$ & $ -0.156$ & $  0.004$ \\
d3t59g35mm00n01 & 5884 &  3.50 & $ -0.00$ & $  0.001$ & $  0.002$ & $  0.051$ \\
d3t59g35mm10n01 & 5890 &  3.50 & $ -1.00$ & $  0.002$ & $  0.003$ & $  0.026$ \\
d3t59g35mm20n01 & 5861 &  3.50 & $ -2.00$ & $ -0.236$ & $ -0.235$ & $  0.035$ \\
d3t59g35mm30n01 & 5873 &  3.50 & $ -3.00$ & $ -0.180$ & $ -0.179$ & $  0.034$ \\
d3t59g40mm00n01 & 5928 &  4.00 & $ -0.00$ & $ -0.077$ & $ -0.077$ & $  0.023$ \\
d3t59g40mm10n02 & 5850 &  4.00 & $ -1.00$ & $ -0.087$ & $ -0.087$ & $  0.017$ \\
d3t59g40mm20n02 & 5856 &  4.00 & $ -2.00$ & $ -0.287$ & $ -0.288$ & $  0.009$ \\
d3t59g40mm30n02 & 5846 &  4.00 & $ -3.00$ & $ -0.233$ & $ -0.232$ & $  0.013$ \\
d3t59g45mm00n01 & 5865 &  4.50 & $ -0.00$ & $ -0.091$ & $ -0.091$ & $  0.019$ \\
d3t59g45mm10n01 & 5923 &  4.50 & $ -1.00$ & $ -0.162$ & $ -0.161$ & $  0.076$ \\
d3t59g45mm20n01 & 5923 &  4.50 & $ -2.00$ & $ -0.285$ & $ -0.284$ & $  0.011$ \\
d3t59g45mm30n01 & 5924 &  4.50 & $ -3.00$ & $ -0.210$ & $ -0.210$ & $  0.011$ \\
d3t63g35mm10n01 & 6210 &  3.50 & $ -1.00$ & $  0.274$ & $  0.274$ & $  0.052$ \\
d3t63g35mm20n01 & 6287 &  3.50 & $ -2.00$ & $  0.019$ & $  0.021$ & $  0.053$ \\
d3t63g35mm30n01 & 6310 &  3.50 & $ -3.00$ & $  0.072$ & $  0.078$ & $  0.048$ \\
d3t63g40mm00n01 & 6229 &  4.00 & $ -0.00$ & $ -0.048$ & $ -0.050$ & $  0.044$ \\
d3t63g40mm10n01 & 6261 &  4.00 & $ -1.00$ & $ -0.081$ & $ -0.081$ & $  0.028$ \\
d3t63g40mm20n01 & 6278 &  4.00 & $ -2.00$ & $ -0.270$ & $ -0.269$ & $  0.017$ \\
d3t63g40mm30n01 & 6269 &  4.00 & $ -3.00$ & $ -0.208$ & $ -0.207$ & $  0.022$ \\
d3t63g45mm00n01 & 6233 &  4.50 & $ -0.00$ & $ -0.110$ & $ -0.110$ & $  0.020$ \\
d3t63g45mm10n01 & 6238 &  4.50 & $ -1.00$ & $ -0.142$ & $ -0.142$ & $  0.014$ \\
d3t63g45mm20n01 & 6323 &  4.50 & $ -2.00$ & $ -0.321$ & $ -0.321$ & $  0.018$ \\
d3t63g45mm30n01 & 6272 &  4.50 & $ -3.00$ & $ -0.230$ & $ -0.230$ & $  0.010$ \\
d3t65g40mm00n01 & 6484 &  4.00 & $ -0.00$ & $  0.014$ & $  0.016$ & $  0.053$ \\
d3t65g40mm10n01 & 6502 &  4.00 & $ -1.00$ & $  0.029$ & $  0.032$ & $  0.042$ \\
d3t65g40mm20n01 & 6534 &  4.00 & $ -2.00$ & $ -0.161$ & $ -0.161$ & $  0.039$ \\
d3t65g40mm30n01 & 6408 &  4.00 & $ -3.00$ & $ -0.163$ & $ -0.163$ & $  0.029$ \\
d3t65g45mm00n01 & 6456 &  4.50 & $ -0.00$ & $ -0.136$ & $ -0.136$ & $  0.039$ \\
d3t65g45mm10n01 & 6456 &  4.50 & $ -1.00$ & $ -0.170$ & $ -0.169$ & $  0.019$ \\
d3t65g45mm20n01 & 6533 &  4.50 & $ -2.00$ & $ -0.316$ & $ -0.316$ & $  0.017$ \\
d3t65g45mm30n01 & 6556 &  4.50 & $ -3.00$ & $ -0.230$ & $ -0.231$ & $  0.023$ \\
d3t68g43mm00n01 & 6726 &  4.25 & $ -0.00$ & $  0.037$ & $  0.041$ & $  0.054$ \\
d3t68g43mm10n01 & 6730 &  4.25 & $ -1.00$ & $  0.020$ & $  0.024$ & $  0.029$ \\
            \hline
\end{longtable}

\end{document}